\documentclass[journal ]{new-aiaa}
\usepackage[utf8]{inputenc}
\usepackage{textcomp}

\usepackage{graphicx}
\usepackage{amsmath}
\usepackage[version=4]{mhchem}
\usepackage{siunitx}
\usepackage{longtable,tabularx}
\title{Comparing design and off-design aerodynamic performance of a natural laminar airfoil}

\author{Aditi Sengupta \footnote{corresponding author, Assistant Professor} and Abhijeet Guha \footnote{M. Tech. Student}}
\affil{Department of Mechanical Engineering, Indian Institute of Technology Dhanbad, Jharkhand, India 826004.}

\begin{document}

\maketitle

\begin{abstract}
Natural laminar flow airfoils are essential technologies designed to reduce drag and significantly enhance aerodynamic performance. A notable example is the SHM1 airfoil, created to meet the requirements of the small-business Honda jet. This airfoil has undergone extensive testing across various operational conditions, including low-speed wind tunnel tests and flight tests across a range of Reynolds numbers and free-stream Mach numbers, as detailed in \lq \lq Natural-laminar-flow airfoil development for a lightweight business jet" by Fujino et al., J. Aircraft, {\bf 40(4)}, 2003. Additionally, investigations into drag-divergence behavior have been conducted using a transonic wind tunnel, with subsequent studies focusing on transonic shock boundary layer interactions through both experimental and numerical approaches. This study employs a series of numerical simulations to analyze the flow physics and aerodynamic performance across different free-stream Mach numbers in the subsonic and transonic regimes. This is achieved by examining computed instantaneous numerical Schlieren for various design conditions (such as low speed, climb, and cruise) and off-design scenarios (including transonic shock emergence, drag-divergence, and shock-induced separation). The dominant time scales, the time-averaged load distributions and boundary layer parameters are compared to provide a comprehensive overview of the SHM1's aerodynamics, establishing benchmark results for optimization of various flow separation and shock control techniques.
\end{abstract}

\section*{Nomenclature}

{\renewcommand\arraystretch{1.0}
\noindent\begin{longtable*}{@{}l @{\quad=\quad} l@{}}
$a$ & speed of sound \\
$c$ & true chord \\
$c_p$ & specific heat at constant pressure \\
$\overline{k}$ & thermal conductivity \\
$p$ & pressure  \\
$u, v$ & streamwise and wall-normal velocities \\
$e_t$ & specific internal energy \\
$J$ & Jacobian of grid transformation \\
$R$ & gas constant \\
$T$ & temperature \\
$C_f$ & skin friction coefficient \\
$C_p$ & coefficient of pressure \\
$U_{\infty}$ & free-stream velocity \\
\multicolumn{2}{@{}l}{Greek letters}\\
$\alpha$ & angle of attack \\
$\gamma$ & adiabatic index \\
$\lambda$ & second coefficient of viscosity \\
$\mu$ & dynamic viscosity \\
$\kappa$ & bulk viscosity \\
$\rho$ & density \\
$\omega$ & vorticity \\
$\xi$, $\eta$ & transformed grid coordinates \\
$\Omega$ & Fourier amplitude of vorticity \\
$\tau_{ii}$ & shear stress components \\
\end{longtable*}}

\section{Introduction}
\label{sec1}

With the growing demand for innovative aircraft and aerial designs to combat climate change, there is an increasing focus on environmentally responsible commercial aviation \cite{coder2014computational}. The primary objectives are to enhance fuel efficiency and reduce aerodynamic drag, which can be achieved by maintaining laminar boundary layer flow. This approach offers a potential tenfold reduction in friction drag compared to turbulent boundary layers \cite{krishnan2017review}. Since friction drag can account for nearly 50\% of the total drag experienced by aircraft during cruise, delaying the transition to turbulence in the boundary layer is crucial for developing fuel-efficient designs \cite{karpuk2024investigation}. Natural laminar flow airfoils present a viable strategy for achieving higher fuel efficiency. Recent studies have demonstrated that integrating natural laminar flow technology with flow separation control \cite{halila2020adjoint} or control of shock waves in the transonic regime \cite{chakraborty2022controlling} can reduce total drag by 15\% or more for typical jetliners at cruise conditions. As a result, natural laminar flow designs, once conceived as strictly experimental and/or conceptual, have gained relevance in modern commercial aircraft, business jets \cite{fujino2003natural}, and unmanned aerial vehicles. This resurgence of interest in natural laminar flow technology can be attributed to recent advancements in high-fidelity simulations that accurately predict laminar-turbulent transitions \cite{halila2019effects} and effectively capture unsteady shock structures \cite{sengupta2021thermal} in transonic aerodynamics. Accurate forecasting of transition onset and shock locations significantly influences boundary layer development, flow separation, friction drag, and maximum lift coefficients \cite{campbell2017building}, all of which are critical to the design and performance of aerodynamic bodies. 

When an aircraft operates outside its intended design conditions, it experiences a phenomenon known as shock boundary layer interaction, where shock waves interact unfavorably with the boundary layer. This is particularly significant in the transonic flow regime, as these off-design events can substantially affect both aerodynamic and thermodynamic properties, altering the flow field in notable ways. Such modifications lead to changes in parameters like pressure distribution and boundary layer characteristics, resulting in increased unsteadiness and higher drag \cite{cole2012transonic}. A thorough investigation of this dynamic phenomenon requires solving the compressible Navier-Stokes equations to accurately capture critical parameters such as shock location and strength, unsteady aerodynamic forces, and potential strategies for reducing the effect of shock waves. While some canonical numerical \cite{larsson2013reynolds} and experimental \cite{barre1996experimental} studies have explored the interaction between shock waves and turbulence, the flow over an airfoil at varying angles of attack presents a more complex scenario due to the presence of variable streamwise pressure gradients, even in laminar flows. Transonic flows are characterized by unsteady shock wave systems. To better understand this intricately time-dependent behavior, researchers have analyzed the effects of downstream periodic pressure perturbations on shock waves \cite{bruce2010experimental}, shedding light on the complex nature of shock-boundary layer interactions.

Previous studies \cite{toure2018numerical, gross2018numerical, quadros2018numerical} on transonic shock boundary layer interactions did not address efforts to modify or control the interactions between shock waves and the underlying boundary layer. Typically, these transient behaviors exhibit large-amplitude normal or near-normal shocks accompanied by low-frequency motion \cite{dussauge2008shock}. Such behavior is problematic due to the resulting unsteady pressure fluctuations on the airfoil, which can lead to increased aerodynamic loads \cite{giannelis2017review}. Another notable aspect of transonic shock boundary layer interactions is the presence of upstream-propagating Kutta waves interacting with the shock system, which includes both oblique and normal shocks \cite{lee2001self}. The low-frequency motion of these shock waves induces similar low-frequency pressure fluctuations on the airfoil surface, a phenomenon referred to as \lq transonic buffeting'. This buffeting can intensify airfoil vibrations and ultimately pose risks of structural failure \cite{giannelis2017review}.

The preceding discussion underscores the necessity for a comprehensive assessment of the aerodynamic performance of a natural laminar flow airfoil under both design and off-design conditions. This assessment will serve as a benchmark for future validation efforts and establish a foundation for strategies aimed at controlling flow separation and shocks to minimize friction drag. Additionally, a thorough investigation into boundary layer characteristics and unsteady separation will contribute to the design of optimized natural laminar flow airfoils. In this study, we will simulate a range of free-stream Mach numbers, Reynolds numbers, and operational conditions for the SHM1 airfoil, which is integral to the design of the Honda business jet \cite{fujino2003natural}. To accurately capture pressure waves, shock structures, and boundary layer interactions, we will employ dispersion relation-preserving compact schemes \cite{sagaut2023global}, which are effective in resolving both temporal and spatial scales within the flow. We ensure the integrity of our numerical approach by implementing an error-free non-overlapping parallelization strategy, which maintains the same level of accuracy as sequential computing \cite{sundaram2023non} through global spectral analysis. This strategy has previously demonstrated its capability in effectively capturing shock boundary layer interactions in compressible transonic flow \cite{chakraborty2022controlling, sengupta2022comparative}, with results validated against flight test data from Fujino \textit{et al.} \cite{fujino2003natural} as well as benchmark wind tunnel results \cite{harris1981two}.

The structure of the paper is as follows: The next section outlines the problem formulation for simulating flow around the SHM1 airfoil, detailing the governing equations and test cases. Section \ref{sec3} describes the numerical methods employed and the validation efforts with experimental data. In Section \ref{sec4}, we present the results and discussions, including instantaneous Schlieren visualizations under various design and off-design conditions, vorticity spectra, and coefficients of pressure and skin friction. Finally, we evaluate the aerodynamic performance by comparing lift and drag coefficients across the different operating conditions. The paper concludes with a summary and final remarks in Section \ref{sec5}.

\section{Problem Formulation of the SHM1 Airfoil}
\label{sec2}

\begin{figure}
\centering
\includegraphics[width=.7\textwidth]{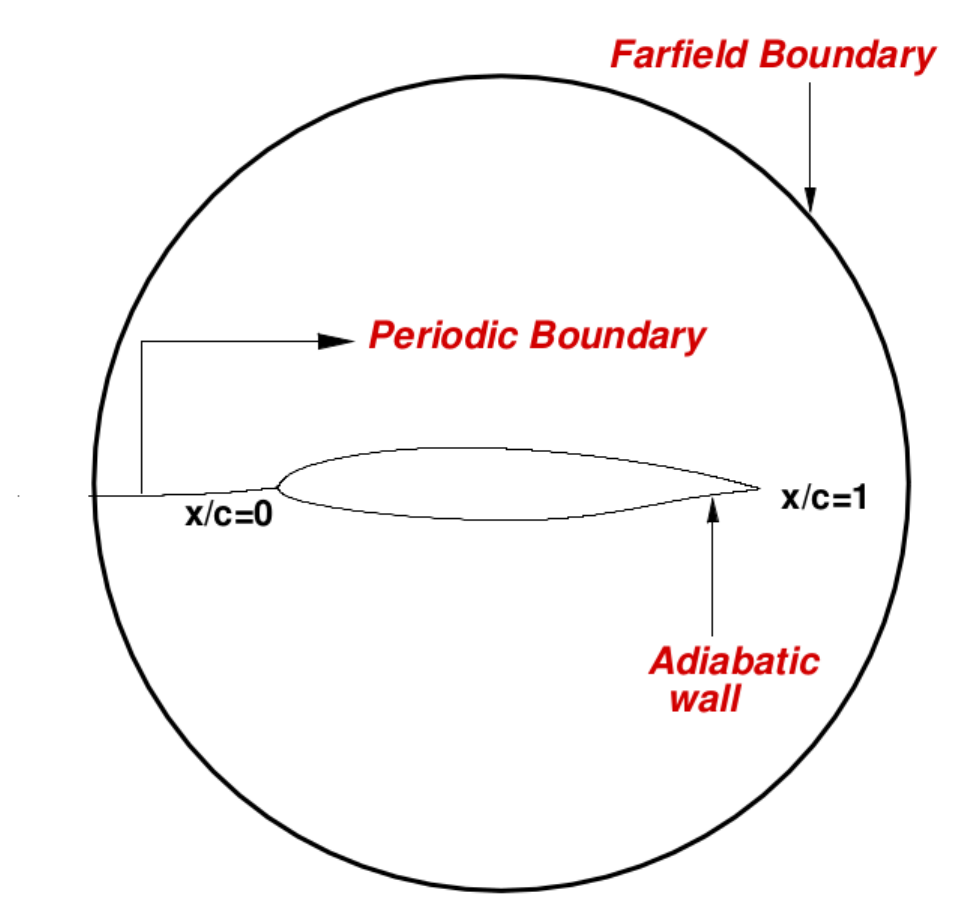}
\caption{Schematic of computational domain and boundary conditions for the natural laminar airfoil, the SHM1.}
\label{fig1}
\end{figure}

Figure \ref{fig1} depicts the schematic of the computational domain, which employs an O-grid topology. This grid was generated using a hyperbolic technique in Pointwise, featuring 1251 points in the azimuthal $(\xi)$ direction and 401 points in the wall-normal $(\eta)$ direction. A cut is introduced along the $\xi$ direction, as shown in Fig. \ref{fig1}, to simplify the domain. This cut allows for natural periodicity of all variables along its boundaries. The outer boundary of the computational domain extends to $16c$, ensuring a nearly uniform grid in the wake region in both directions. A shock-free mesh is crucial for accurately capturing the dynamics of traveling waves in compressible flow around airfoils. The grid points are concentrated near the wall in the $\eta$ direction, as well as near the leading edge, trailing edge, and the region ahead of the separation point on the upper surface in the $\xi$ direction. The numerical simulations solve the two-dimensional compressible Navier-Stokes equations, formulated by established notations, as detailed in previous works \cite{sengupta2022comparative, sengupta2024thermal}.

\begin{eqnarray}
\frac{\partial {\hat{Q}}}{\partial t} + \frac{\partial {\hat{E}}}{\partial x} + \frac{\partial {\hat{F}}}{\partial y} = \frac{\partial {\hat{E_{v}}}}{\partial x} + \frac{\partial {\hat{F_{v}}}}{\partial y}
\label{eq10} 
\end{eqnarray}

\noindent where the conservative variables are given as, $\hat{Q} = [ \rho \;\; \rho u \;\; \rho v \;\; e_t]^T$. The convective flux vectors are similarly given as,

$\hat{E} = [ \rho u \;\; \rho u^2 +p \;\; \rho uv \;\; (\rho e_t + p) u]^T$
 
$\hat{F} = [ \rho v \;\; \rho uv \;\; \rho v^2 +p \;\; (\rho e_t + p) v]^T$

\noindent and the viscous flux vectors are given as,

$\hat{E_{v}}= [ 0\;\; \tau _{xx} \;\; \tau _{xy} \;\; (u \tau _{xx} +v \tau _{xy} -q _{x})]^T$

$\hat{F_{v}}= [0 \;\; \tau _{yx} \;\; \tau _{yy} \;\; (u \tau _{yx} +v \tau _{yy} -q _{y})]^T$ 

In the given expressions, $\rho$, $e_{t}$, and $p$ denote dimensionless values of density, total specific energy, and pressure, respectively. These physical variables are normalized with respect to the free-stream density ($\rho_{\infty}$), free-stream velocity ($U_{\infty}= \frac{M_{s}}{\sqrt{\gamma RT_{\infty}}}$), free-stream temperature ($T_{\infty}$), free-stream dynamic viscosity ($\mu_{\infty}$), length scale ($c$), and the time scale, ($c/U_{\infty}$). Here, $\gamma=1.4$ represents the specific heat capacity ratio or the adiabatic index for air. The dimensionless parameters, namely the Prandtl number ($Pr$), free-stream Reynolds number ($Re_{\infty}$), and free-stream Mach number, $M_{s}$, are defined as follows:

\begin{equation*}
Pr = \frac{\mu C_{p}}{\overline{k}}	;\ Re_{\infty} = \frac{\rho_{\infty} U_{\infty} c}{\mu_{\infty}}	;\ M_{s} = \frac{U_{\infty}}{a_{\infty}}
\end{equation*}    

\noindent where $a_{\infty}$ is the free-stream speed of sound. The heat conduction terms involved are given by

\begin{eqnarray*}
q_{x}= - \frac{\mu}{Pr Re_{\infty} (\gamma -1) M_{s}^2} \frac{\partial {T}}{\partial x}\\
q_{y}= - \frac{\mu}{Pr Re_{\infty} (\gamma -1) M_{s}^2} \frac{\partial {T}}{\partial y}
\end{eqnarray*}

\noindent The components of the symmetric Newtonian viscous stress tensors, $\tau_{xx},\tau_{xy},\tau_{yx},\tau_{yy}$, are defined as 

\begin{equation*}
\tau_{xx} = \frac{1}{Re_{\infty}}\Big[\Big( \frac{4}{3}\mu + \frac{\kappa}{\mu_{\infty}}\Big)\frac{\partial u}{\partial x} + \Big(-\frac{2}{3}\mu + \frac{\kappa}{\mu_{\infty}}\Big)\frac{\partial v}{\partial y}  \Big]
\end{equation*}
\begin{equation*}
\tau_{yy} = \frac{1}{Re_{\infty}}\Big[\Big( \frac{4}{3}\mu + \frac{\kappa}{\mu_{\infty}}\Big)\frac{\partial v}{\partial y} + \Big(-\frac{2}{3}\mu + \frac{\kappa}{\mu_{\infty}}\Big)\frac{\partial u}{\partial x}  \Big]
\end{equation*}
\begin{equation*}
\tau_{xy} = \tau_{yx} = \frac{\mu}{Re_{\infty}}\Big[\frac{\partial u}{\partial y} + \frac{\partial v}{\partial x} \Big]
\end{equation*}

\noindent The equations from the Cartesian space ($x$, $y$) are transformed to body-fitted computational grid ($\xi, \eta$) using the following relations: $\xi = \xi (x, y)$ and $\eta = \eta (x, y)$. The transformed plane equations in strong conservation form are given as, 

\begin{eqnarray}
\frac{\partial {Q}}{\partial t} + \frac{\partial {E}}{\partial \xi} + \frac{\partial {F}}{\partial \eta} = \frac{\partial {E_{v}}}{\partial \xi} + \frac{\partial {F_{v}}}{\partial \eta}
\label{eq11} 
\end{eqnarray}   

\noindent with the state variables and flux vectors, given as 
\begin{eqnarray*}
Q &=& \hat{Q}/J\\
E &=& (\xi _{x} \hat{E} +\xi _{y} \hat{F})/J\\
F &=& (\eta _{x} \hat{E} +\eta _{y} \hat{F})/J\\
E_{v} &=& (\xi _{x} \hat{E_{v}} +\xi _{y} \hat{F_{v}})/J\\
F_{v} &=& (\eta _{x} \hat{E_{v}} +\eta _{y} \hat{F_{v}})/J
\end{eqnarray*}

\noindent Here $J$ is the Jacobian of the grid transformation given by

\begin{equation*}
J=\frac{1}{x_{\xi} y_{\eta} -x_{\eta} y_{\xi}}
\end{equation*}

\noindent The grid metrics, $\xi_{x}$, $\xi_{y}$, $\eta_{x}$, and $\eta_{y}$, are computed during the creation of the O-grid using a hyperbolic grid generation technique in Pointwise. These are expressed as follows: $\xi_{x} = J y_{\eta}$; $\xi_{y} = -J x_{\eta}$; $\eta_{x} = -J y_{\xi}$; $\eta_{y} = J x_{\xi}$. This grid transformation ensures that the solid airfoil boundary aligns with one of the grid lines ($\eta = 0$). Using these transformations, the heat conduction terms in the transformed plane are given by,

\begin{eqnarray*}
q_{x}= -\frac{\mu}{Pr Re_{\infty} (\gamma -1) M_{s}^2} (\xi_{x} T_{\xi}+\eta_{x} T_{\eta})\\
q_{y}= -\frac{\mu}{Pr Re_{\infty} (\gamma -1) M_{s}^2} (\xi_{y} T_{\xi}+\eta_{y} T_{\eta})
\end{eqnarray*}

\noindent The viscous stress components in the transformed plane are given by

\begin{equation*}
\tau_{xx} = \frac{1}{Re_{\infty}}\Big[\Big( \frac{4}{3}\mu + \frac{\kappa}{\mu_{\infty}}\Big)\Big(\xi_{x} u_{\xi} + \eta_{x} u_{\eta}\Big) - 
 \Big(-\frac{2}{3}\mu + \frac{\kappa}{\mu_{\infty}}\Big)\Big(\xi_{y} v_{\xi} + \eta_{y} v_{\eta}\Big) \Big] \; 
\end{equation*}

\begin{equation*}
\tau_{yy} = \frac{1}{Re_{\infty}}\Big[\Big( \frac{4}{3}\mu + \frac{\kappa}{\mu_{\infty}}\Big)\Big(\xi_{y} v_{\xi} + \eta_{y} v_{\eta}\Big) - 
 \Big(-\frac{2}{3}\mu + \frac{\kappa}{\mu_{\infty}}\Big)\Big(\xi_{x} v_{\xi} + \eta_{x} u_{\eta}\Big) \Big]
\end{equation*}

\begin{equation*}
\tau_{xy} = \tau_{yx} = \frac{\mu}{Re_{\infty}}\Big[\xi_{y} u_{\xi} + \eta_{y} u_{\eta} + \xi_{x} v_{\xi} + \eta_{x} v_{\eta} \Big]
\end{equation*}

On the airfoil surface, a no-slip wall boundary condition is applied to the velocity components, specified as $u = v = 0$. An adiabatic wall condition is enforced on the heat conduction terms, ensuring no heat transfer occurs across the airfoil surface: $(q_{x})_{wall} = (q_{y})_{wall} = 0$. In the far field, the flow is approximated as one-dimensional in the $\eta$-direction, where a non-reflective characteristic-like boundary condition \cite{pulliam1986implicit} derived from the Euler equations is implemented. These conditions are based on the signs of the eigenvalues of the linearized one-dimensional Euler equations, which depend on whether the flow is subsonic or supersonic at the far field. At the domain's inflow, characteristic variables are set to the free-stream values corresponding to the specified Mach number. At the outflow, variables are extrapolated from the interior when the local Mach number is supersonic. To prevent spurious acoustic wave reflections that could compromise the physical integrity of the domain and to ensure accurate implementation of the far-field boundary conditions, the outer boundary is positioned at approximately $16c$.

\section{Numerical Methodology and Validation}
\label{sec3}

The current simulations utilize highly accurate dispersion relation preserving compact schemes for the spatial discretization and time integration of the governing equations \cite{sagaut2023global}. Convective flux derivatives are computed using an optimized upwind compact scheme, $OUCS3$, with explicit boundary closures \cite{sagaut2023global}. To maintain the isotropic nature of viscous flux derivatives, a second-order central difference method (CD2) is employed for discretization. A novel parallelization strategy \cite{sundaram2022non} is implemented without overlap points at the sub-domain boundaries, minimizing errors associated with parallelization by computing derivatives using the interior compact scheme with global spectral analysis. Time integration is conducted using a fourth-order, four-stage Runge-Kutta method (RK4) with a time step of $2.5 \times 10^{-6}$ for all the reported simulations, to ensure all spatial and temporal scales in the flow are captured. This numerical framework has been validated in a prior work \cite{sundaram2023non} by comparing the pressure distribution on the SHM1 airfoil with experimental results from Fujino {\it et al.} \cite{fujino2003natural} for the following non-dimensional parameters: $Re_{\infty} = 13.6 \times 10^{6}$, $M_{s} = 0.62$ and $\alpha = 0.27^{\circ}$. Additionally, the same numerical methods were used \cite{sengupta2013direct} to validate against benchmark wind tunnel results reported by Harris \cite{harris1981two} for the NACA0012 airfoil. The large wind tunnel employed \cite{harris1981two} mitigated issues associated with testing in transonic wind tunnels \cite{garbaruk2003numerical}, such as wall interference and three-dimensional effects.

The study includes six test cases, detailed in Table \ref{tab1}, each initiated with free-stream conditions relevant to various design and off-design scenarios for the SHM1 airfoil \cite{fujino2003natural}. The free-stream temperature and density are set as: $T_{\infty}= 288.15 K$ and $\rho _{\infty}= 1.2256 kg/m^{3}$. Case-1 represents conditions for testing the SHM1 in a low-speed wind tunnel. Case-2 corresponds to a climb condition with $M_s = 0.31$, where low profile drag is desirable. Case-3 reflects a cruise condition for the SHM1 at $M_s = 0.62$. Cases 4 through 6 operate in the transonic regime, where shock boundary layer interactions occur. In Case-4, a small normal shock is observed in the Schlieren visualizations \cite{sengupta2021thermal}. Case-5 simulates the drag divergence Mach number, resulting in a strong normal shock on the suction surface \cite{fujino2003natural}. Conversely, Case-6 features transonic shock boundary layer interactions that induce separation on the suction surface, leading to a $\Lambda$-shock and wedge-shaped shock structure \cite{sengupta2024thermal}. The $Re_{\infty}$ and angles of attack, $\alpha$ are selected to replicate conditions from experimental low-speed and transonic wind tunnel tests, as well as flight test data \cite{fujino2003natural}.

\begin{table}[h!]
\centering
\caption{Numerical parameters used and description of the test cases reported.}
\vspace{1mm}
\begin{tabular}{|c| c| c| c| c|}
\hline
 Case & $M_s$ & $Re_{\infty}$ & $\alpha$  \\ [0.5ex]
 \hline\hline
Case-1: Low speed condition & 0.134 & $4.8 \times 10^{6}$ & $0.27^{\circ}$ \\
Case-2: Climb condition & 0.310 & $13.6 \times 10^{6}$ & $0.27^{\circ}$  \\
Case-3: Cruise (design) condition & 0.620 & $13.6 \times 10^{6}$ & $0.27^{\circ}$  \\
Case-4: Transonic shock condition & 0.720 & $16.2 \times 10^{6}$ & $0.38^{\circ}$  \\
Case-5: Drag-divergence condition & 0.730 & $8 \times 10^{6}$  & $0.50^{\circ}$ \\
Case-6: Shock-induced separation & 0.780 & $8 \times 10^{6}$  & $0.50^{\circ}$  \\ [1ex]
 \hline
\end{tabular}

\label{tab1}
\end{table} 

\begin{figure}
\centering
\includegraphics[width=.8\textwidth]{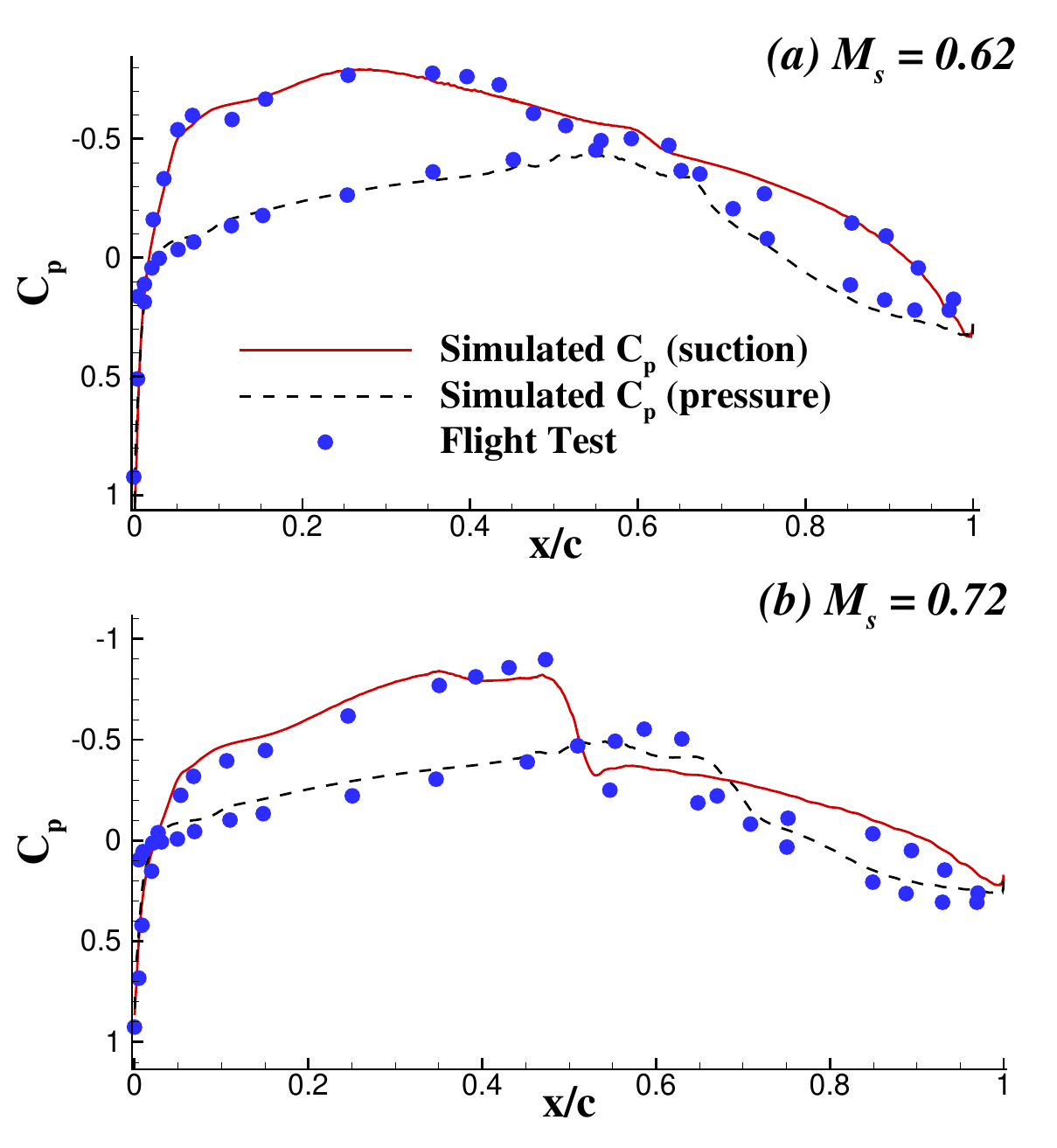}
\caption{Comparison of simulated time-averaged $C_p$ for (a) $M_s = 0.62$ and (b) $M_s = 0.72$ with Fujino {\it et al.} \cite{fujino2003natural}'s flight test data. The period of time-averaging is from $t = 20$ to 100 in increments of 0.05.}
\label{fig2}
\end{figure}

Figure \ref{fig2} compares the numerically computed time-averaged coefficient of pressure, $C_p$ distributions for a design condition ($M_s$ = 0.62) and an off-design condition ($M_s = 0.72$) with experimental flight test data from Fujino et al. \cite{fujino2003natural}. The results show good agreement between the simulated and experimental $C_p$ on both the suction and pressure surfaces. The $C_p$ distribution reveals a plateau, characteristic of natural laminar airfoils \cite{somers1992subsonic}. The SHM1 is designed to maintain a favorable pressure gradient on the suction surface up to $x/c = 0.42$, followed by a concave pressure recovery, which is seen in Fig. \ref{fig2}(a) for $M_s = 0.62$. In contrast, the pressure surface exhibits a favorable pressure gradient up to $x/c = 0.63$ to enhance drag reduction \cite{fujino2003natural}. A steeper pressure recovery is evident on the pressure surface under the design condition, as shown in Fig. \ref{fig2}(a). The simulation also accurately captures the shock location, as depicted in Fig. \ref{fig2}(b), occurring at $x/c \approx 0.45$ on the suction surface \cite{sengupta2021thermal} for the off-design transonic condition ($M_s = 0.72$) \cite{sengupta2021thermal}. The steep pressure rise induced by the shock is frequently observed upstream of the point where the shock interacts with the suction surface, as the shock transmits a \lq \lq pressure signal" in the upstream direction, within the subsonic inner part of the boundary layer \cite{delery1986shock}.

\section{Results and Discussion}
\label{sec4}

In this section, we compare the numerical Schlieren for the different operating conditions outlined in Table \ref{tab1}, emphasizing the varying flow physics and associated features. We also discuss the dominant time scales for these various design and off-design conditions, along with the aerodynamic performance of the SHM1 airfoil. These insights serve as supplementary benchmark datasets to the experimental data provided by Fujino et al. \cite{fujino2003natural}.

\subsection{Comparing low speed, climb and cruise (design) conditions with transonic shock boundary layer interactions (off-design)}

\begin{figure}
\centering
\includegraphics[width=.85\textwidth]{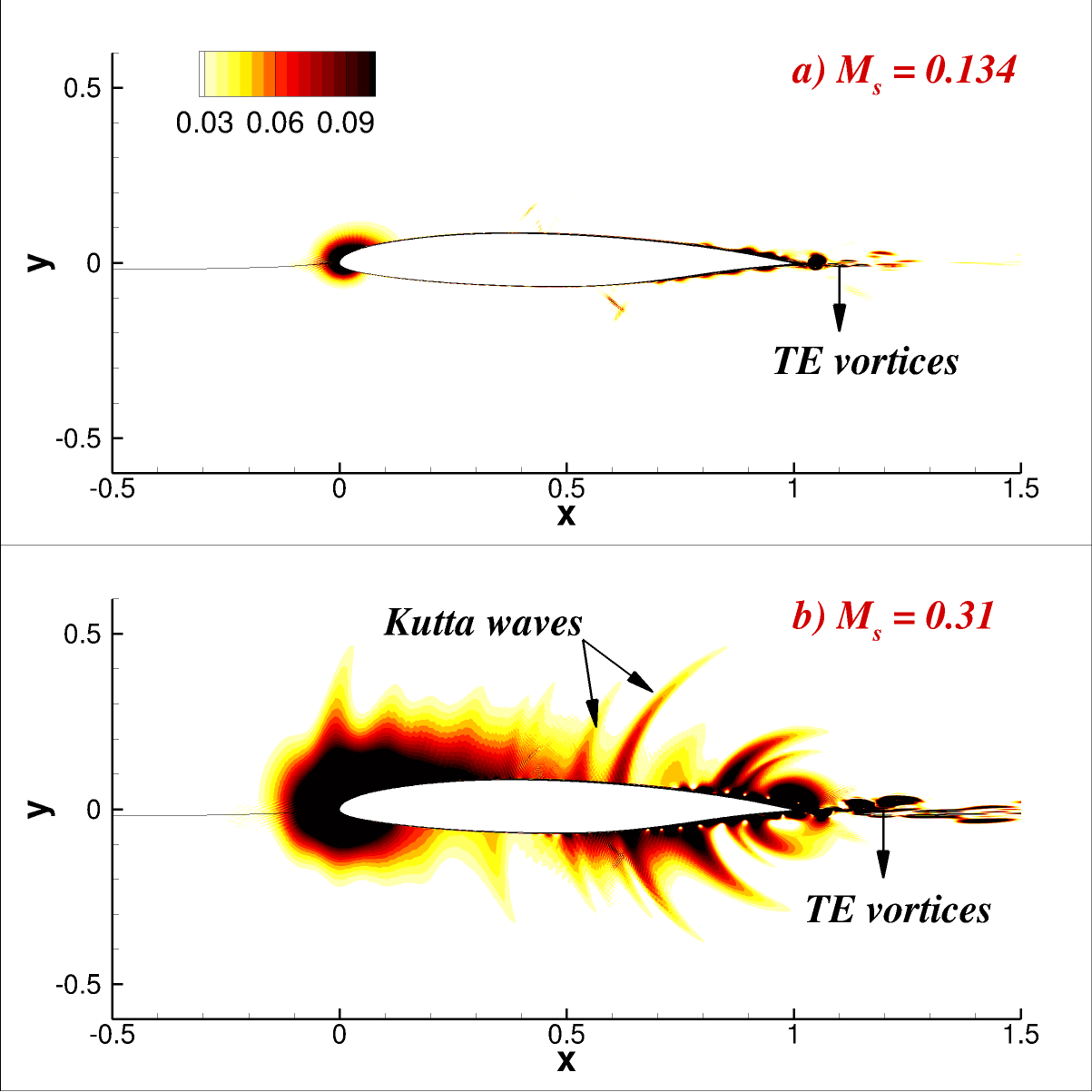}
\caption{Comparison of numerical Schlieren contours for low speed operation at (a) $M_s = 0.134$ and for climb condition at (b) $M_s = 0.31$.}
\label{fig3}
\end{figure}

\begin{figure}
\centering
\includegraphics[width=.85\textwidth]{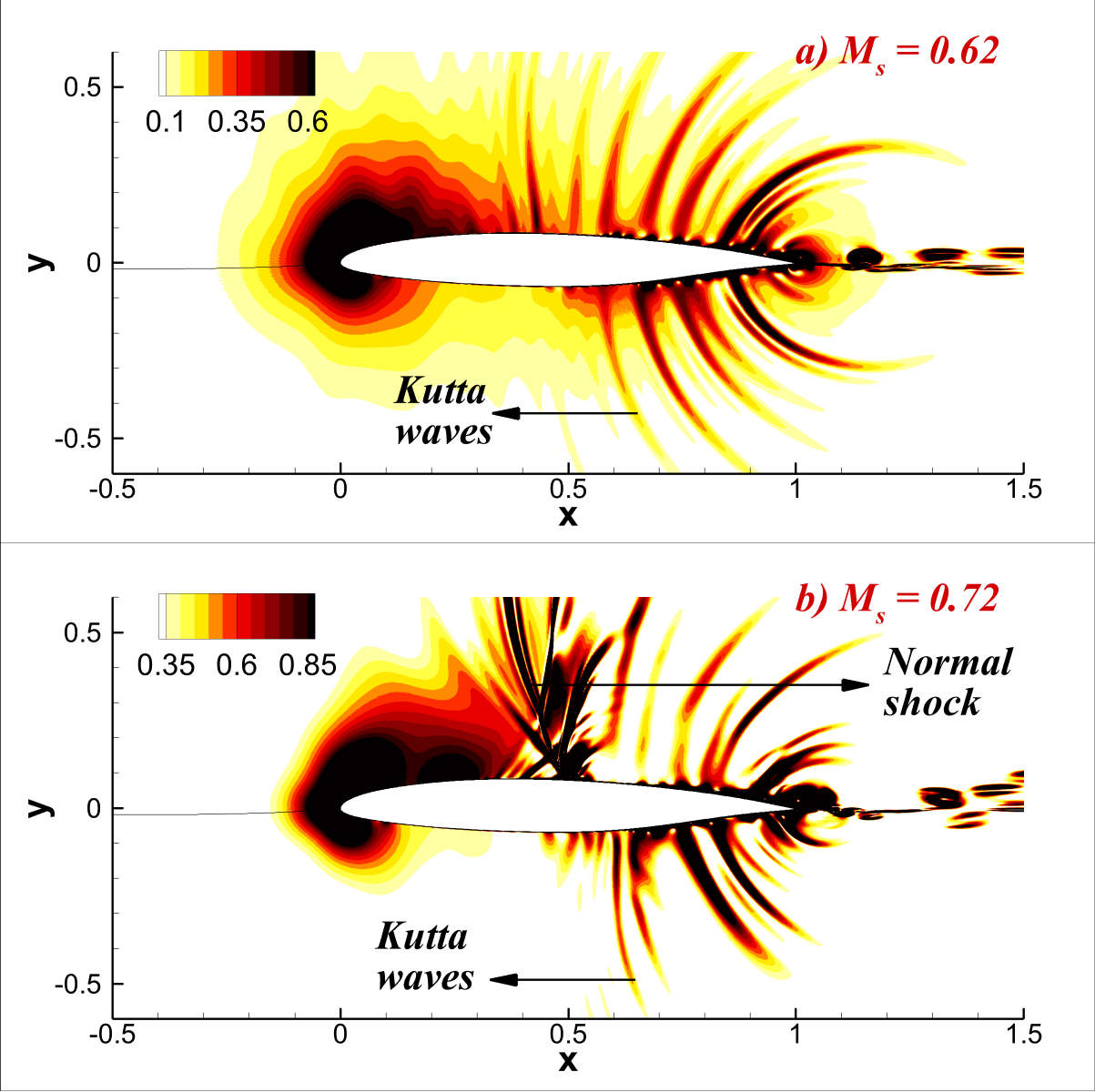}
\caption{Comparison of numerical Schlieren contours for design condition at (a) $M_s = 0.62$ and for transonic shock appearance at (b) $M_s = 0.72$, respectively.}
\label{fig4}
\end{figure}

\begin{figure}
\centering
\includegraphics[width=.85\textwidth]{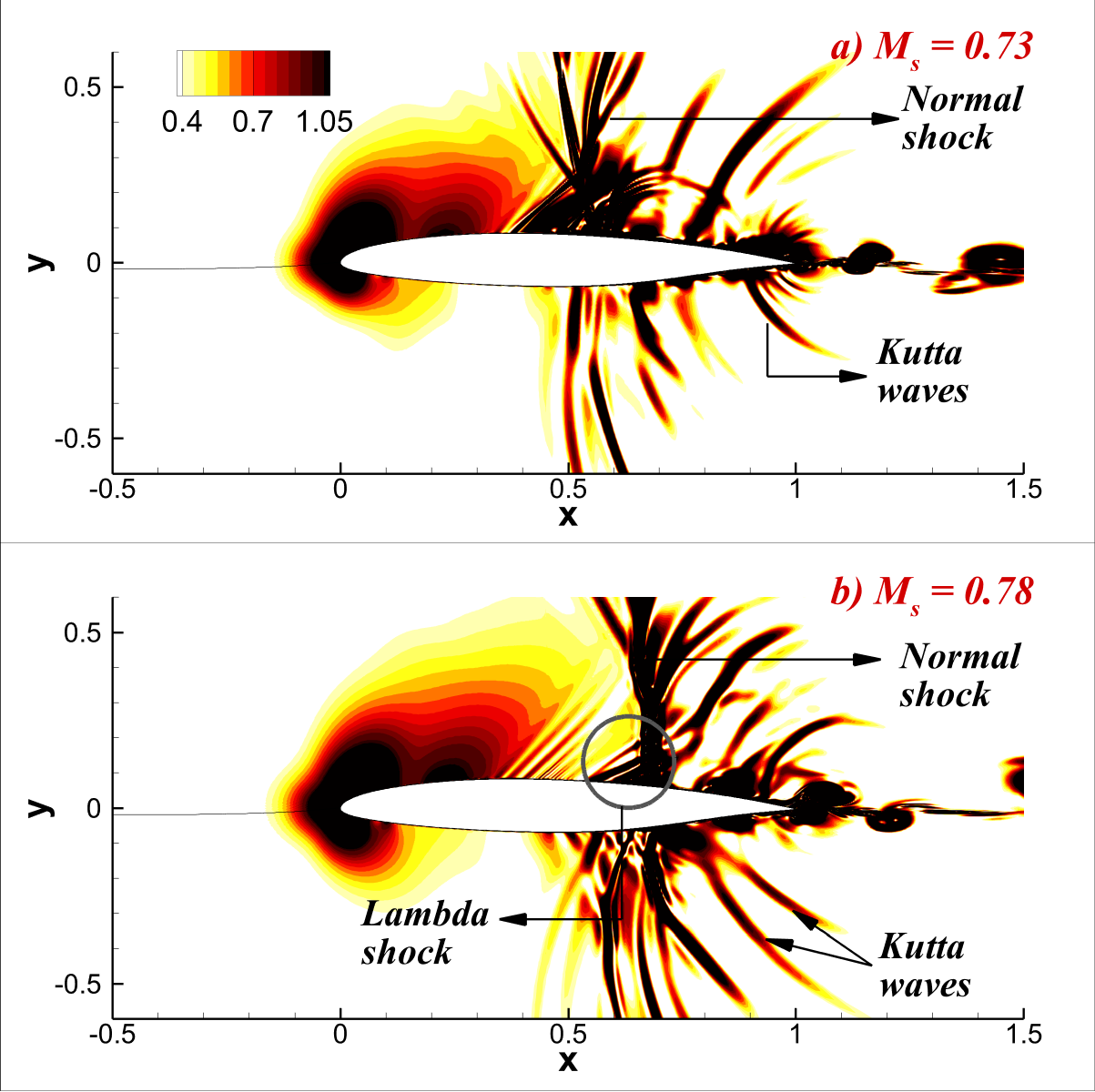}
\caption{Comparison of numerical Schlieren contours for drag divergence condition at (a) $M_s = 0.73$ and for shock-induced separation at (b) $M_s = 0.78$, respectively.}
\label{fig5}
\end{figure}

Following the design requirements of the Honda jet, the SHM1 airfoil has been optimized to operate for the design conditions of Cases 1-3 of Table \ref{tab1}. Transonic operation and the ensuing shock boundary layer interactions are investigated using Cases 4-6 of Table \ref{tab1}. The flow features for these various operating conditions are explored via the instantaneous numerical Schlieren contours in Figs. \ref{fig3} to \ref{fig5}. 

Visualization of flow features in experiments with the Schlieren technique depict optical effects using density gradients. This has been extended in creating numerical Schlieren \cite{samtaney2000visualization} where the authors numerically plotted $\nabla \rho$ in the domain using Robert’s edge detection technique. The density derivatives were evaluated following a low order accuracy discrete formula. Here, $\nabla \rho$ is calculated by
calculating the $\xi$ and $\eta$ derivatives of $\rho$ using the high accuracy
compact schemes \cite{sagaut2023global}. In Fig. \ref{fig3}, we compare the numerical Schlieren evaluated at $t = 50$ for the low speed ($M_s = 0.134$) and climb ($M_s = 0.31$) conditions. For the SHM1, the leading edge of the airfoil has been designed to induce transition near itself, as noted for both $M_s$ simulated here. This, in turn, removes the problem of loss in lift due to contamination near the leading edge \cite{fujino2003natural}. Similarly, near the aft portion of the airfoil, the SHM1 design typically induces separation which aids in reducing the pitching moment. For the low speed operation in Fig. \ref{fig3}(a), vorticity is shed from the trailing edge following Helmholtz's theorem \cite{houghton2003aerodynamics}. As the $M_s$ is increased in Fig. \ref{fig3}(b) to 0.31 (climb condition), in addition to the trailing edge vortices, upstream propagating pressure waves are noted. When the flow accelerates, density gradients are induced normal to the surface of the airfoil which induce pressure pulses at regular intervals near the trailing edge \cite{tijdeman1977investigations}. In the inviscid part of the flow, these acoustic waves propagate along paths away from the airfoil following the Kutta condition \cite{houghton2003aerodynamics}. These upstream propagating pressure waves are termed as Kutta waves \cite{lee2001self}. While propagating in the upstream direction, Kutta waves exhibit nonlinear interactions with one another, enhancing in strength. 

In Fig.\ref{fig4}, we compare the numerical Schlieren corresponding to two completely varied flight conditions: for $M_s = 0.62$ in Fig. \ref{fig4}(a), a cruise condition is simulated which is a \lq shock-free' flow. This $M_s$ is near the upper limit without shock appearance, where little to no wave drag is expected. On the other hand, for $M_s = 0.72$, depicted in Fig. \ref{fig4}(b), we simulate a case with transonic shock boundary layer interactions which has negative implications for the profile drag. The cruise condition visualized in Fig. \ref{fig4}(a) shows the presence of trailing edge vortices and symmetric Kutta waves on pressure and suction surfaces of the airfoil. Compared to the climb condition in Fig. \ref{fig3}(b), the nonlinear interactions between the Kutta waves are stronger and are spread over a longer extent of the airfoil, particularly on the suction surface. Due to the interactions between the Kutta waves and the boundary layer, the flow is more chaotic and a multi-periodic signal is expected. For the SHM1, beyond a critical $M_s$ of 0.70, somewhere in the flow the local Mach number becomes unity. For $M_s = 0.72$, depicted in Fig. \ref{fig4}(b), a supersonic region appears along the suction surface, which is terminated by a normal shock wave. Through this shock wave, the flow velocity is reduced from supersonic to subsonic locally. Interestingly, the shock wave centred at $x/c \approx 0.45$, although termed as a normal shock is not perfectly normal to the airfoil surface. Only the foot of the shock is normal to the airfoil surface, the remaining portion is curved forward. This can be explained by the requirement of wall-normal velocity upstream and downstream of the shock to decrease, along a convex contour \cite{zierep2003new}. Such a requirement is not compatible with a completely straight shock.   

The transonic shock boundary layer interactions on the SHM1 are compared using the flow visualizations in Fig. \ref{fig5} for two operational regimes: at the drag divergence Mach number in Fig.\ref{fig5}(a) and during shock-induced separation for $M_s = 0.78$ in Fig. \ref{fig5}(b). At the elevated free-stream Mach number of $M_s = 0.73$, the normal shock observed for $M_s = 0.72$ in Fig. \ref{fig4}(b), moves backward to $x/c \approx 0.5$ while both shock strength and size of the supersonic region increases. A weak shock structure appears on the pressure surface also. The spread of the supersonic regime disrupts the upstream propagation of the Kutta waves leading to strong localised perturbation waves away from the airfoil surface. Multiple time periods (associated with the transonic shock boundary layer interactions) are expected in the spectrum of the associated signal. When the pressure jump through the shock wave has become sufficiently large, shock-induced separation of the turbulent boundary layer occurs, as in Fig. \ref{fig5}(b). The local Mach numbers just upstream of the shock wave are in the range 1.25-1.3. Strong normal shocks appear on both suction and pressure surface, as a consequence. Here, in addition to the normal shock, a \lq $\Lambda$-shock' has been identified in the numerical Schlieren. As the supersonic flow passes a concave corner, an oblique shock wave occurs which changes the direction of the flow. This oblique shock formation is followed by the near-normal shock wave. These merge to form the wedge-shaped \lq $\Lambda$-shock' along the suction surface. Here too, the transonic shock boundary layer interactions are delayed due to heightened $M_s$ to be centred around $x/c \approx 0.6$, concurrent with prior experiments and simulations \cite{fujino2003natural, sengupta2024thermal} of $M_s = 0.78$. 

\begin{figure}
\centering
\includegraphics[width=.85\textwidth]{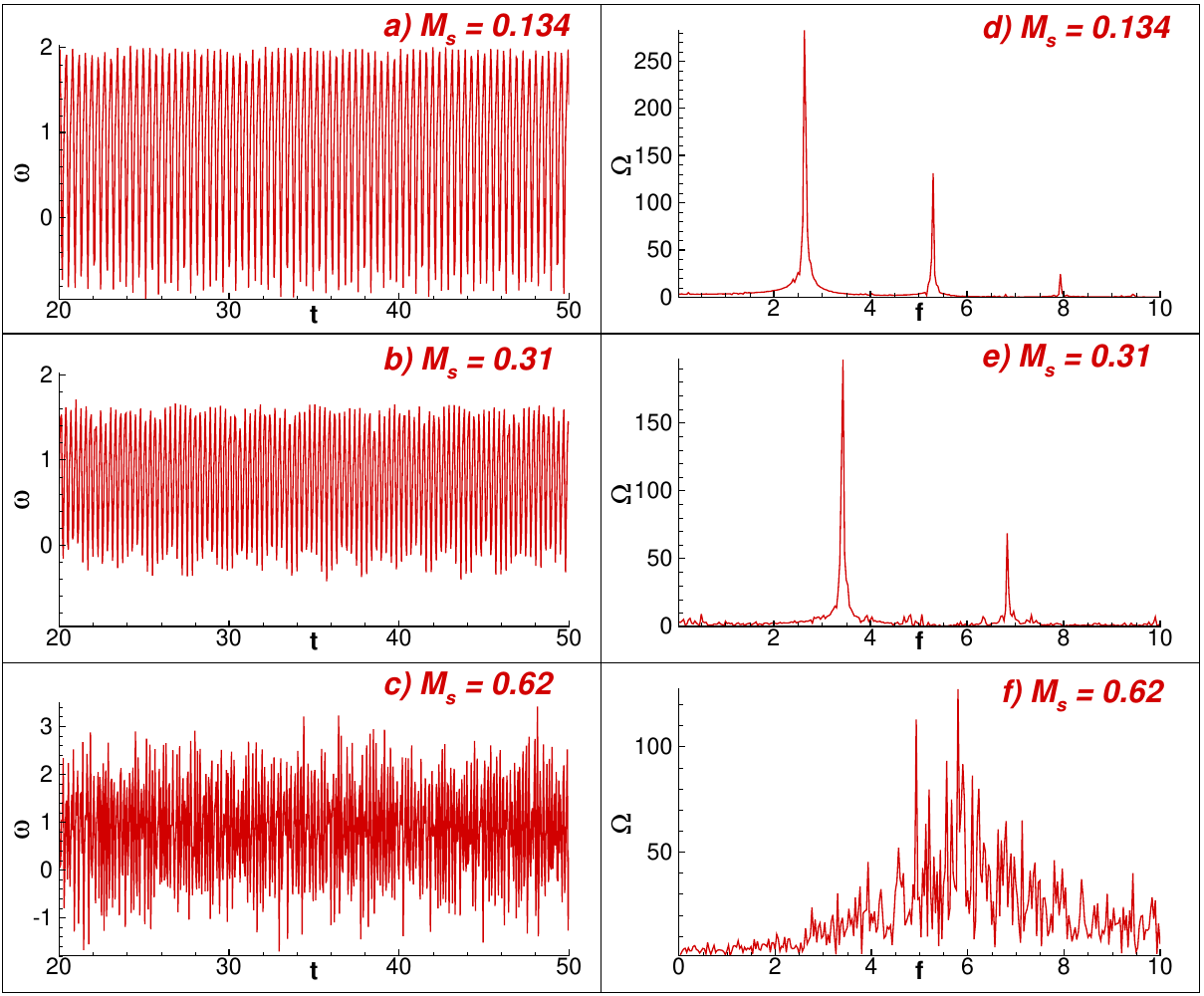}
\caption{Time-series of vorticity probed near the trailing edge of the suction surface for (a) $M_s = 0.134$, (b) $M_s = 0.31$ and (c) $M_s = 0.62$. The corresponding FFT are shown in frames (d), (e), and (f).}
\label{fig6}
\end{figure}

Fig.\ref{fig6} compares the time-series of vorticity in frames (a to c) for low speed, climb, and cruise conditions, respectively. The probe location is along the aft portion of the suction surface for these design conditions of the SHM1 airfoil. The corresponding spectra are computed by performing Fast Fourier transform (FFT) of the time series in frames (d to f) for Cases 1-3 in Table \ref{tab1}. For the low speed operation at $M_s = 0.134$, the time-series in Fig. \ref{fig6}(a) reveals the presence of a singular time-period, affirmed by the spectrum in Fig. \ref{fig6}(b). A dominant peak is noted at $f = 2.85Hz$, followed by a sub-dominant peak at a superharmonic frequency of $f = 5.6Hz$. These are sub-harmonics of the Kelvin-Helmholtz shedding frequency, $f \approx 14-16 Hz$ \cite{sengupta2024separation}, associated with the vortex shedding near the trailing edge for $M_s = 0.134$ in Fig. \ref{fig3}(a). For the climb condition in Fig. \ref{fig6}(b), the time-series shows the presence of multiple time periods. The spectrum in Fig. \ref{fig6}(b) corroborates this in the form of multiple peaks of insignificant amplitude across the frequency plane. The dominant and sub-dominant peaks are at $f = 3.5Hz$ and its superharmonic, i.e. $f = 7Hz$, respectively. These are still in the range of the Kelvin-Helmotz shedding frequency. The altered time-period is due to interactions of the inviscid Kutta waves with the trailing edge vortices for this case, as seen in Fig. \ref{fig3}(b). With highly nonlinear interactions among the Kutta waves and with the trailing edge vortices at $M_s = 0.62$ (as seen in Fig. \ref{fig4}(a)), the time-series and associated spectrum in Figs. \ref{fig6}(c) and (f), demonstrate a chaotic, multi-periodic nature of the flow. No discernible peaks can be ascertained in the frequency plane. Interestingly, the vorticity magnitude is the highest for $M_s = 0.62$, however due to redistribution of the Fourier amplitude of vorticity across various frequencies, the maximum amplitude is noted for the lowest $M_s = 0.134$, where a preferential dominant frequency could be ascribed to the vortex shedding phenomenon. 

\begin{figure}
\centering
\includegraphics[width=.85\textwidth]{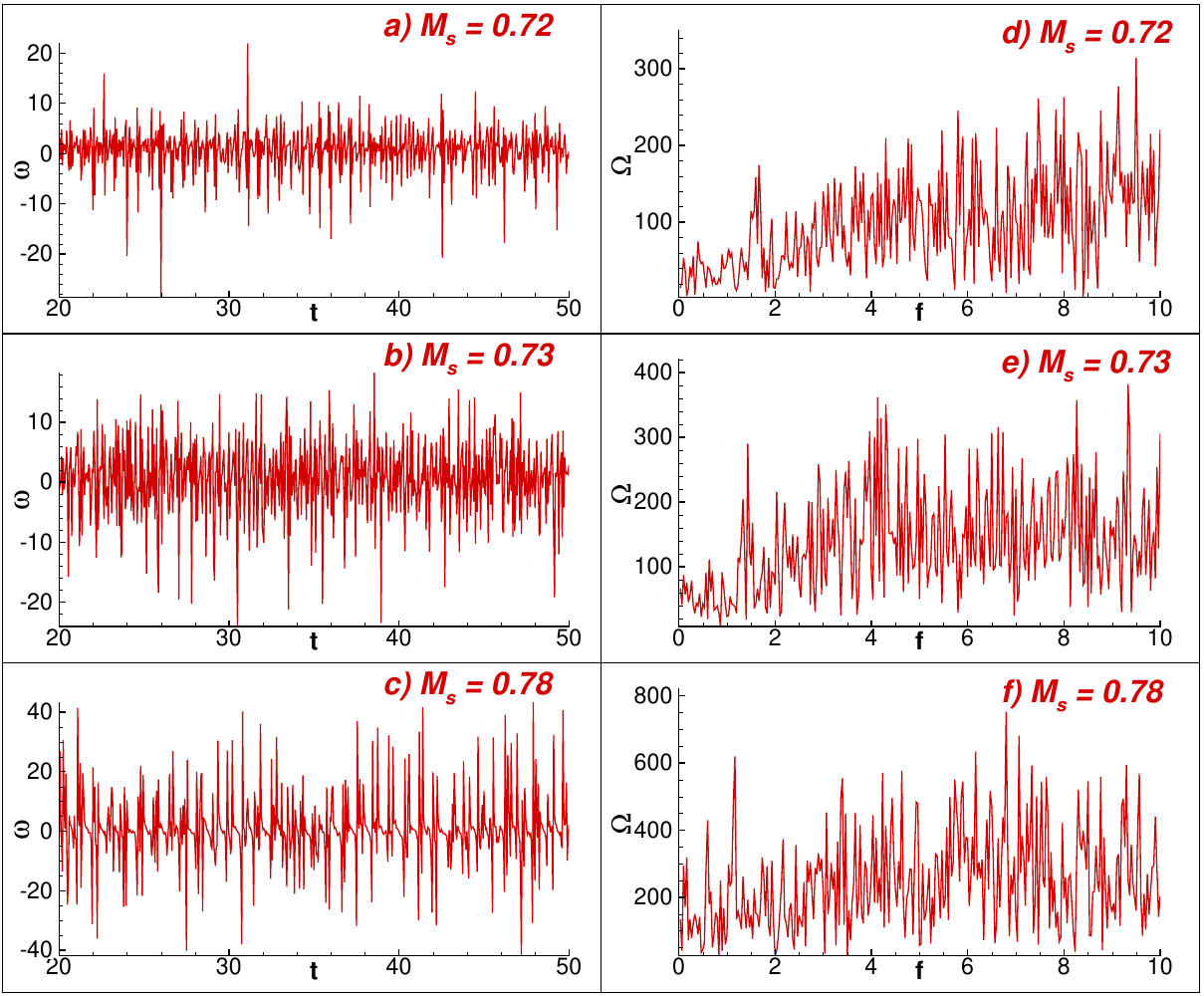}
\caption{Time-series of vorticity probed near the trailing edge of the suction surface for (a) $M_s = 0.72$, (b) $M_s = 0.73$ and (c) $M_s = 0.78$. The corresponding FFT are shown in frames (d), (e), and (f).}
\label{fig7}
\end{figure}

Fig.\ref{fig7} compares the time-series of vorticity in frames (a to c) for the first appearance of transonic shock boundary layer interactions, drag-divergence operation and shock-induced separation conditions, respectively. The probe location is along the aft portion of the suction surface for these off-design conditions of the SHM1 airfoil. The corresponding spectra of the time series are shown in frames (d to f) for Cases 4-6 of Table \ref{tab1}. After the appearance of the normal shock, apart from nonlinear interactions among Kutta waves and with the boundary layer, transonic shock boundary layer interactions evoke multiple time-periods in the flow. This is evidenced by the chaotic multi-periodic nature of the spectra in Figs. \ref{fig7}(d)-(f). Compared to the design conditions in Fig. \ref{fig6}, the magnitude of vorticity generated is higher. The associated Fourier amplitude is also greater due to the transonic shock boundary layer interactions. Among these off-design conditions also, as the free-stream $M_s$ is increased, the Fourier amplitude increases with vorticity redistribution across the entire frequency range. However, for the drag-divergence $M_s = 0.73$ in Fig. \ref{fig7}(b), the vorticity generation is distributed over a wider range of temporal scales, due to the transition from the normal shock to the shock-induced separation in case of $M_s = 0.78$. Contrary to the design condition, however, no discernible peak is noted in the frequency plane due to Kelvin-Helmholtz vortex shedding or otherwise.

The multi-periodic nature of the flow with elevated free-stream $M_s$ indicates a time-dependent variation in pressure and shear forces. To investigate the temporal evolution of aerodynamic loads resulting from these unsteady forces acting on the airfoil surface, we need to analyze the computed results from the time-averaged flow. The interactions with the boundary layer and unsteady separations has to be explored further. We also compute coefficients of lift, $C_l$, and drag, $C_d$, by integrating static pressure and shear forces acting over the control surface.

\subsection{Characterization of the boundary layer and aerodynamic performance}

In this section, we examine the time-averaged flow field for design and off-design conditions of the SHM1 airfoil, described in Table \ref{tab1}. This exploration involves the computation of coefficients of pressure and skin friction. Furthermore, the integrated $C_l$ and $C_d$ and aerodynamic efficiency are tabulated for the various computed cases to provide insights into the aerodynamic performance of the SHM1, initialized with different free-stream $M_s$.

\begin{figure}
\centering
\includegraphics[width=.85\textwidth]{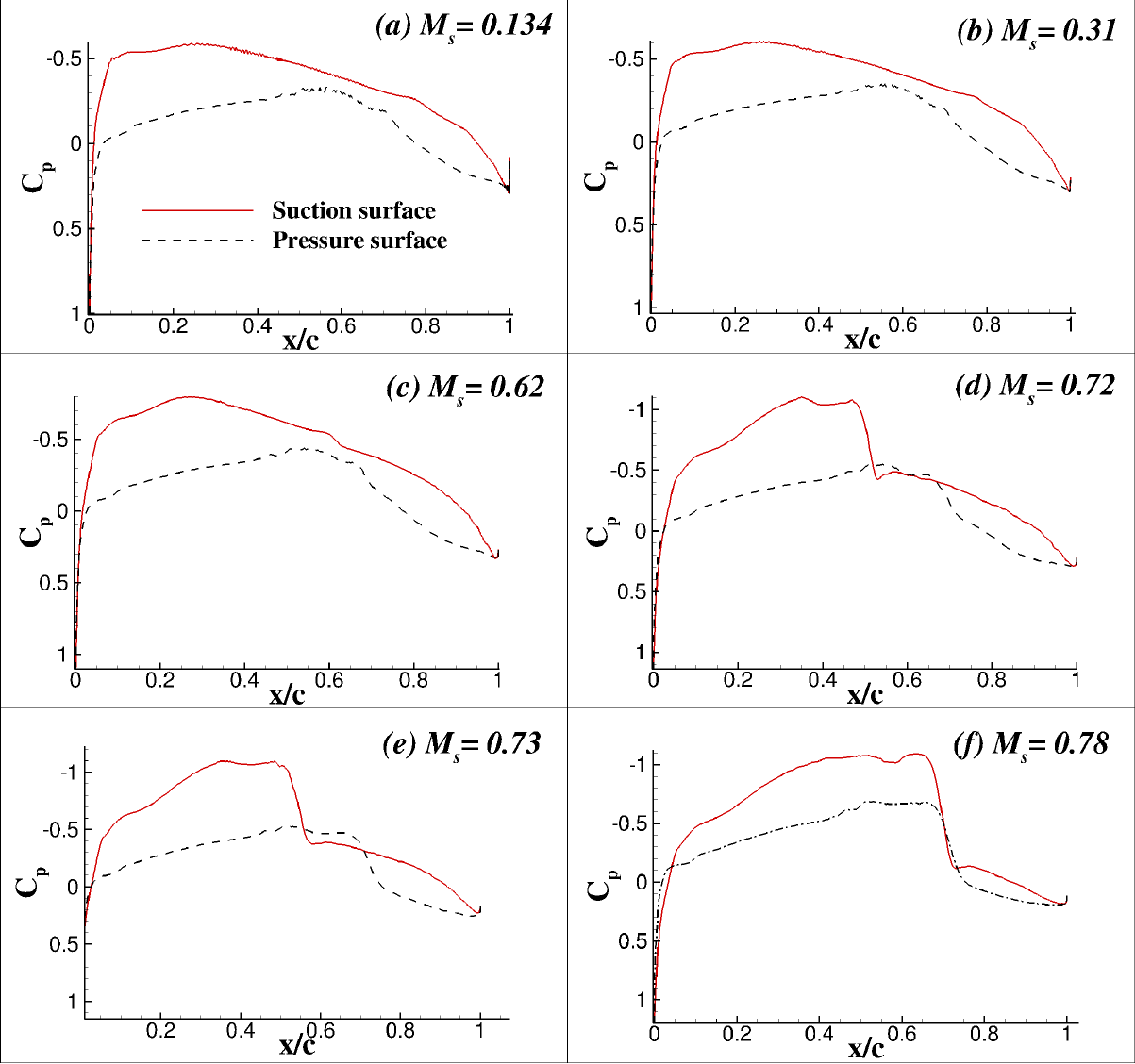}
\caption{Streamwise variation of time-averaged $C_p$ for (a) $M_s = 0.134$, (b) $M_s = 0.31$, (c) $M_s = 0.62$, (d) $M_s = 0.72$, (e) $M_s = 0.73$, and (f) $M_s = 0.78$; on the suction and pressure surfaces. The period of time-averaging is from $t = 20$ to 100 in intervals of 0.05.}
\label{fig8}
\end{figure}

Fig.\ref{fig8} shows the time-averaged streamwise $C_p$ distribution for both the design and off-design conditions of the SHM1 airfoil, tabulated in Table \ref{tab1}. For the low speed, climb, and cruise conditions shown in frames (a) to (c), the pressure distribution on the suction and pressure surfaces are similar. A minor kink in the $C_p$ distribution on the suction surface for $M_s = 0.62$, for $0.45 < x/c < 0.55$ due to the nonlinear interactions among the Kutta waves and with the underlying boundary layer. This region is found to be susceptible to shock formation on further increase of the $M_s$. For all computed $M_s$, (prior to the shock location for off-design conditions), a flat plateau is noted in the pressure distribution on the suction surface, a typical characteristic of natural laminar airfoils \cite{selig1995natural}. On the pressure surface, for all $M_s$, there is a concave pressure recovery after $x/c = 0.63$, following the design constraints of the SHM1 airfoil \cite{fujino2003natural}. For $M_s = 0.73$ and 0.78, a pressure spike is observed on the pressure surface in addition to the suction surface. This spike is steeper for $M_s = 0.78$ with shock-induced separation evoking strong normal shocks on the lower surface of the airfoil. For the off-design conditions shown in Figs. \ref{fig8}(d)-(f), a pressure spike is observed at the approximate shock location. For $M_s = 0.72$, 0.73, and 0.78, this spike starts from $x/c = 0.45$, 0.5 and 0.6, respectively, which is the approximate central location of the shock structures observed in Figs. \ref{fig4} and \ref{fig5}. A steeper spike indicates a higher gradient of variables across the shock structure. With increase in free-stream $M_s$, the separation and shock formation is delayed. The steepest spike is observed for $M_s = 0.78$, which demonstrated the largest separated region and shock extent in the numerical Schlieren of Fig. \ref{fig5}. 

\begin{figure}
\centering
\includegraphics[width=.8\textwidth]{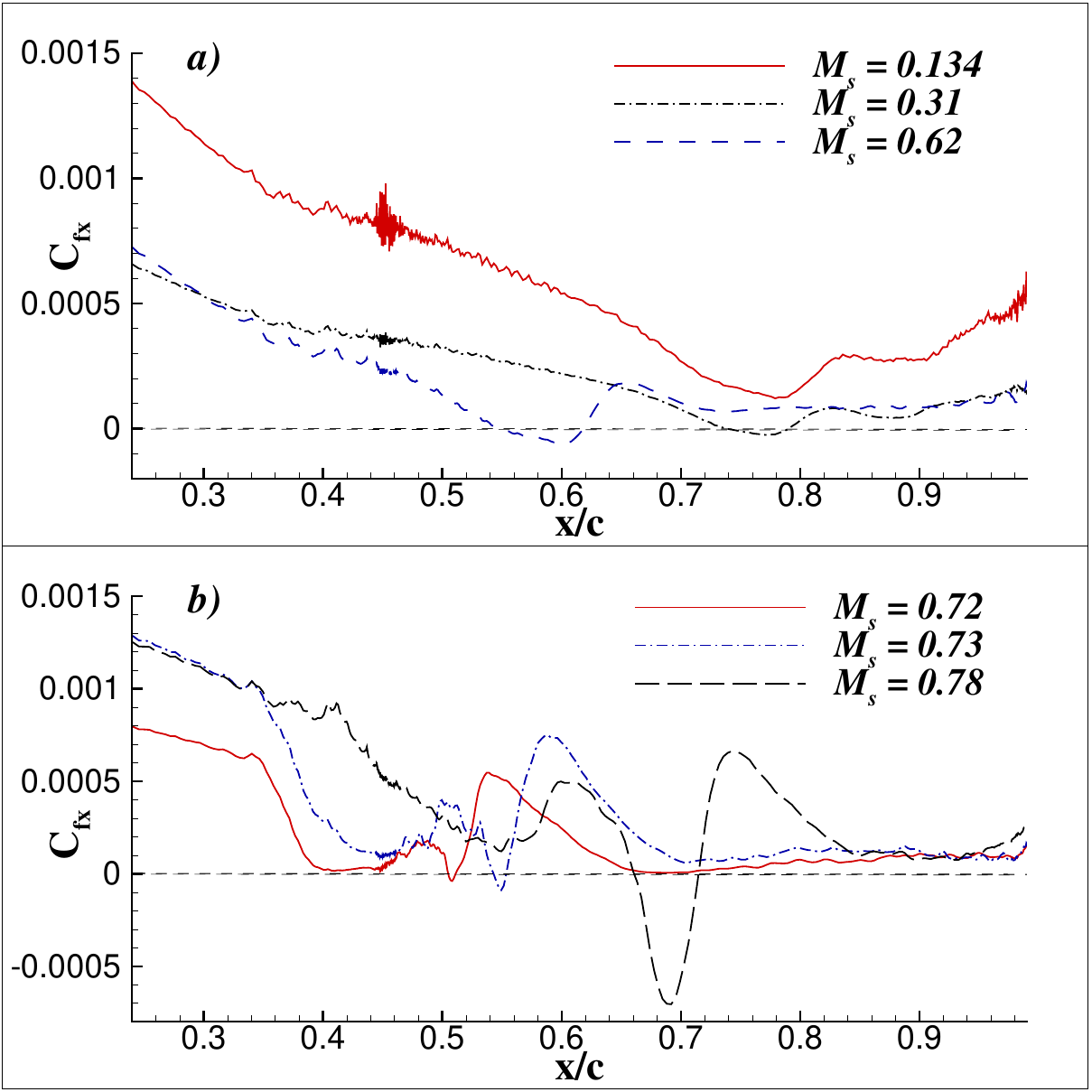}
\caption{Streamwise variation of time-averaged $C_{fx}$ for (a) $M_s = 0.134$, (b) $M_s = 0.31$, (c) $M_s = 0.62$, (d) $M_s = 0.72$, (e) $M_s = 0.73$, and (f) $M_s = 0.78$; on the suction surface. The period of time-averaging is from $t = 20$ to 100 in intervals of 0.05.}
\label{fig9}
\end{figure}

In Fig.\ref{fig9}, time averaged streamwise variation of skin-friction coefficient $C_{fx}$ is compared for design conditions in frame (a) and off-design conditions in frame (b). The initial zero crossing of the $C_{fx}$ line provides an approximate location for flow separation while the second crossing is the approximate flow reattachment location. In Fig. \ref{fig9}(a), the red solid line ($M_s = 0.134$), does not cross the zero line indicating no separation for this free-stream Mach number. For the climb condition, ($M_s = 0.31$), a very short separation bubble is noted near the aft portion of the suction surface. A similar separation event is observed at the cruise condition, i.e. $M_s = 0.62$ at $x/c = 0.6$. The earlier separation can be explained by the nonlinear interaction between the Kutta wave and the boundary layer. The off-design cases in Fig. \ref{fig9}(b) show the presence of small separation bubbles at the approximate shock locations for $M_s = 0.72$ and 0.73. Separation is delayed with elevated free-stream $M_s$, with the largest separation bubble observed at $x/c = 0.68$ for $M_s = 0.78$. The shock-induced separation shows the largest streamwise extent of the separated region with largest skin friction drag.   

Table \ref{tab2} quantifies the aerodynamic performance of the SHM1 airfoil in its design and off-design conditions. From the table, one can conclude that the least drag is observed for Case-6 (shock-induced separation) and highest for Case-5 (drag-divergence). The climb condition (Case-2) also shows an elevated $C_d$ compared to the low speed and cruise conditions due to the elevated thrust requirement. The highest lift is generated for the drag-divergence condition (Case-5), but since it is offset by the largest $C_d$, the overall aerodynamic efficiency is low in this case, compared to the cruise condition, which has understandably the best $C_l/C_d$. For Case-6, despite having the lowest $C_d$, it has a poor aerodynamic performance due to the loss in lift during the shock-induced separation on the suction surface, compared to the other transonic operation points (Cases-3 to 5). 

\begin{table}[h!]
\centering
\caption{Computed mean coefficients of lift, drag, and aerodynamic efficiency for flow over the SHM-1 airfoil. Time-averaging is done from $t = 20$ to 100.}
\vspace{1mm}
\begin{tabular}{|c| c| c| c| c|}
\hline
 Case & $\Bar{C_l}$ & $\Bar{C_d}$ & $\Bar{C_l}/\Bar{C_d}$  \\ [0.5ex]
 \hline\hline
Case-1: $M_s = 0.134$ & 0.25089 & $4.66969 \times 10^{-3}$ & $5.37273$ \\
Case-2: $M_s = 0.31$ & 0.24585 & $8.15143 \times 10^{-3}$ & $30.15998$  \\
Case-3: $M_s = 0.62$ & 0.30302 & $3.05563 \times 10^{-3}$ & $99.16864$  \\
Case-4: $M_s = 0.72$ & 0.31284 & $5.13265 \times 10^{-3}$ & $60.95032$  \\
Case-5: $M_s = 0.73$ & 0.32190 & $8.68036 \times 10^{-3}$  & $37.08479$ \\
Case-6: $M_s = 0.78$ & 0.26214 & $2.56901 \times 10^{-3}$  & $10.20380$  \\ [1ex]
 \hline
\end{tabular}

\label{tab2}
\end{table}

\section{Summary and Conclusions}
\label{sec5}

The present study investigates the aerodynamic performance of the SHM1 airfoil (depicted in Fig. \ref{fig1}) across various design and off-design conditions, with a focus on understanding flow physics and associated dominant time-scales. For the off-design cases, in particular, the effects of transonic shock boundary layer interactions as a function of increasing free-stream Mach number is also explored. Utilizing dispersion relation preserving highly accurate compact schemes for discretization of the governing compressible Navier-Stokes equations, six implicit large eddy simulations are performed for a range of free-stream Mach numbers. The numerics are validated against the experiments of Fujino {\it et al.} \cite{fujino2003natural}, revealing a good agreement of the computed pressure distribution with that in the flight test in Fig. \ref{fig2}. The instantaneous flow features, spectra, pressure distributions, skin friction coefficients, and overall aerodynamic efficiency provide critical insights into the operation of the SHM1 for the following design scenarios: (i) low speed operation, (ii) climb conditions, (iii) cruise conditions; and the following off-design operation points: (i) first appearance of the near-normal transonic shock, (ii) drag-divergence condition, and (iii) shock-induced separation. 

The various flow features are traced via density gradients in the numerical Schlieren plots in Figs. \ref{fig3} to \ref{fig5}. For low-speed operation, vortex shedding from the trailing edge follows the characteristic shedding frequency of Kelvin-Helmholtz eddies \cite{sengupta2024separation}. Upon increasing the free-stream Mach number to that in the climb condition, in addition to the trailing edge vortices, upstream propagating pressure pulses are noted in the inviscid part of the flow, termed as Kutta waves \cite{lee2001self}. These undergo nonlinear interaction with each other and the underlying boundary layer at the cruise condition, leading to a perfect symmetric arrangement on the pressure and suction surfaces of the SHM1 airfoil. For a free-stream Mach number of 0.72, a near-normal shock \cite{zierep2003new} is obtained on the suction surface which locally alters the supersonic flow to a subsonic one. At the drag-divergence Mach number, a stronger normal shock with a larger spread across the suction surface is observed. A weaker shock is obtained on the pressure surface also. For the shock-induced separation, $\Lambda$ shock waves are obtained due to merging of the normal shock waves with the oblique ones on the suction surface. Strong, nonlinear shock waves are obtained on the pressure surface, which are expected to contribute detrimentally to the skin friction drag at this operational regime. 

The dominant time-scales in the flow are identified by examining the spectrum of vorticity for the six computed cases in Figs. \ref{fig6} and \ref{fig7}. While the low-speed and climb conditions reveal distinct peaks in the frequency plane at sub-harmonics of the Kelvin-Helmholtz shedding frequency, the transonic shock boundary layer interactions evoke multi-periodicity in the form of normal, oblique and $\Lambda$ shocks. The associated spectra are chaotic with no discernible peak along the frequency plane. The time-averaged streamwise pressure coefficient computed in Fig. \ref{fig8} demonstrate characteristic plateaus along the suction surface indicative of natural laminar airfoils. Pressure spikes were identified at shock locations for off-design conditions along both suction and pressure surfaces (for free-stream Mach numbers of 0.73 and 0.78). A concave pressure recovery is noted on the pressure surface beyond $x/c = 0.63$, matching the design requirements of the SHM1 \cite{fujino2003natural}. The steepest spikes in the pressure distribution are noted for the shock-induced separation due to the presence of stronger shocks spread over a wider section of the suction surface. A delayed separation, with stronger shocks are noted with an increase in the free-stream Mach number. The analysis of skin-friction coefficients in Fig. \ref{fig9} revealed flow separation behavior, with shorter separation bubbles at lower Mach numbers and larger, delayed separation regions as free-stream Mach numbers increased. These findings underscore the complex interactions between Kutta waves, the boundary layer and the shock structures.

The aerodynamic performance of the SHM1 airfoil was quantified in Table \ref{tab2} by computing the integrated coefficients of lift and drag. The ratio of lift and drag coefficient is a metric for the aerodynamic efficiency, which has also been compared for the six test cases. It has been highlighted that Case-6 (shock-induced separation) has the least drag but also suffers from a significant loss of lift. Conversely, Case-5 (drag-divergence) exhibited the highest lift, although this was offset by substantial drag, resulting in low aerodynamic efficiency. The cruise condition emerged as the most aerodynamically efficient, achieving the best ratio of $C_l$ to $C_d$. 

In conclusion, this work enhances our understanding of the SHM1 airfoil's performance across different operating conditions. The results provide valuable benchmarks for future optimization efforts, particularly in managing shock-induced separations and enhancing overall aerodynamic efficiency in transonic flight regimes. The insights gained from this study are essential for improving the design and operational strategies of lightweight business jets and other aircraft utilizing similar airfoil configurations.

\section*{Acknowledgments}
The authors acknowledge the use of the high-performance computing facility, ARYABHATA at Indian Institute of Technology Dhanbad for computing all the cases reported here.

\bibliography{aiaa_shm1}

\end{document}